\documentclass[a4paper,11pt,reqno]{amsart}
\usepackage{amsmath,amsfonts,amssymb,enumerate}

\def\p{\partial}

\def\R{\mathbb{R}}

\begin{document}

\title{DIFFERENTIAL ALGEBRAS ON $\kappa$-MINKOWSKI SPACE\\AND ACTION OF THE LORENTZ ALGEBRA}
\author{STJEPAN MELJANAC}
\address{Rudjer Bo\v{s}kovi\'{c} Institute, Bijeni\v{c}ka cesta b.b., 10000 Zagreb, Croatia}
\email{meljanac@irb.hr}

\author{SA\v{S}A KRE\v{S}I\'{C}-JURI\'{C}}
\address{Faculty of Natural and Mathematical Sciences, University of Split, Teslina 12, 21000 Split, Croatia}
\email{skresic@pmfst.hr}

\author{RINA \v{S}TRAJN}
\address{Rudjer Bo\v{s}kovi\'{c} Institute, Bijeni\v{c}ka cesta b.b., 10000 Zagreb, Croatia}
\email{rina.strajn@gmail.com}

\date{}

\begin{abstract}
We propose two families of differential algebras of classical dimension on $\kappa$-Minkowski space. The algebras are constructed
using realizations of the generators as formal power series in a Weyl super-algebra. We also propose a novel realization of
the Lorentz algebra $\mathfrak{so}(1,n-1)$ in terms of Grassmann-type variables. Using this realization we construct an action
of $\mathfrak{so}(1,n-1)$ on the two families of algebras. Restriction of the action to $\kappa$-Minkowski space is covariant.
In contrast to the standard approach the action is not Lorentz covariant except on constant one-forms, but it does not require
an extra cotangent direction.
\end{abstract}
\maketitle

\keywords{Keywords: $\kappa$-Minkowski space, Lorentz algebra,realizations,differential algebra,\\ covariance}

\subjclass{PACS numbers: 02.20.Sv, 02.20.Uw, 02.40.Gh}

%%%%%%%%%%%%%%%%%%%%%%%%%%%%%%%%%%%%%%%%%%%%%%%%%%%%%%%%%%%%%%%%%%%%%%%%%%%%%%%%%%%%%%%%%%%%%%%%%%%%%%%%%%%%%%%%%%%%%%%%%%%%%%%%%

\section{Introduction}

Noncommutative (NC) geometry has been proposed for many years as a
suitable model for unification of quantum field theory and
gravity. Noncommutative spaces have been studied from many
different points of view, including operator theory \cite{Connes}
and Hopf algebras \cite{Majid-1}--\cite{Agostini-2}. In particular,
the notion of differential calculus on NC spaces has been studied
in Refs. \cite{Woronowicz}--\cite{Meljanac-2}. It is known that
many classes of NC spaces do not admit differential calculi of
classical dimensions which are fully covariant under the expected
group of symmetries \cite{Beggs}. This quantum anomaly for
differential structures is usually fixed by introducing extra
cotangent directions.

In this paper we focus our attention to $\kappa$-Minkowski space.
This is a Lie algebra type NC space which appears as a deformation
of ordinary Minkowski space-time within the framework of doubly
special relativity
(DSR) \cite{Amelino-Camelia-1}--\cite{Daskiewicz}. The symmetry
algebra for DSR is obtained by deforming the ordinary Poincar\'{e}
algebra into a Hopf algebra known as $\kappa$-Poincar\'{e}
algebra \cite{Kowalski-Glikman-1}--\cite{Kowalski-Glikman-3}.
Different bases of $\kappa$-Poincar\'{e} algebra correspond to
different versions of DSR theory \cite{Kowalski-Glikman-1}. The
$\kappa$-deformed Poincar\'{e} algebra as deformed symmetry of the
$\kappa$-Minkowski space-time inspired many authors to construct
quantum field theories (see e.g. Refs. \cite{Daskiewicz}--\cite{mstw11}) and electrodynamics on
$\kappa$-Minkowski space-time \cite{hjm11}--\cite{dj11}, or to
modify particle statistics \cite{Govindarajan-1}, \cite{Govindarajan-2}.

Bicovariant differential calculus on $\kappa$-Minkowski space-time
was considered by Sitarz in Ref. \cite{Sitarz}. He has shown that
if the bicovariant calculus is required to be Lorentz covariant,
then one obtains a contradiction with a Jacobi identity for the
generators of the differential algebra. This contradiction is
resolved by adding an extra cotangent direction (one-form) which
has no classical analogue. Thus, the differential calculus in 3+1
dimensions developed in Ref. \cite{Sitarz} is five-dimensional.
This work was generalized to $n$ dimensions by Gonera et al. in
Ref. \cite{Gonera}. There have been several attempts to deal with
this issue in $\kappa$-Euclidean and $\kappa$-Minkowski
spaces \cite{Dimitrijevic}--\cite{Meljanac-2}. In Ref.
\cite{Bu-Kim} Bu et al. constructed a differential algebra on
$\kappa$-Minkowski space from Jordanian twist of the Weyl algebra
and showed that the algebra is closed in four dimensions. In their
approach they extended the $\kappa$-Poincar\'{e} algebra with a
dilatation operator and used a coproduct of the Lorentz generators
which is different from the one used in Ref. \cite{Sitarz}. In
Refs. \cite{Meljanac-1} and \cite{Meljanac-2} differential
algebras of classical dimension on $\kappa$-Euclidean and
$\kappa$-Minkowski spaces are constructed. In this approach
one-forms are obtained from an action of a deformed exterior
derivative on NC coordinates. Different deformations of exterior
derivative and NC coordinates lead to different versions of
differential calculus compatible with $\kappa$-deformation.

In the present work we propose new families of differential algebras (denoted $\mathfrak{D}_1$ and $\mathfrak{D}_2$) on $\kappa$-Minkowski
space $\mathcal{M}_\kappa$ using realizations of the generators as formal power series in a Weyl super-algebra. We also present a novel realization
of the Lorentz algebra $\mathfrak{so}(1,n-1)$ in terms of Grassmann-type variables. This realization is used to define an action of
$\mathfrak{so}(1,n-1)$ on $\mathfrak{D}_1$ and $\mathfrak{D}_2$ which is consistent with commutation relations in
$\mathfrak{D}_1$ and $\mathfrak{D}_2$ and Lorentz covariant on constant one-forms.
Restriction of the action to $\mathcal{M}_\kappa$ is covariant, thus $\mathcal{M}_\kappa$ is an
$\mathfrak{so}(1,n-1)$-module algebra.

The paper is organized as follows. In Sec. 2 we discuss briefly a
method for constructing differential algebras based on
realizations of NC coordinates and exterior derivative as formal
power series in a Weyl super-algebra. We construct two families of
differential algebras  $\mathfrak{D}_1$ and $\mathfrak{D}_2$ of
classical dimension on $\kappa$-Minkowski space, and discuss their
properties. We also show that by the same method the differential
algebras in Refs. \cite{Sitarz} and \cite{Majid-2} can be
constructed. In Sec. 3 we propose an action of
$\mathfrak{so}(1,n-1)$ on $\mathfrak{D}_1$ and $\mathfrak{D}_2$
using a realization of $\mathfrak{so}(1,n-1)$ in terms of
Grassmann-type variables. The action is not Lorentz covariant
except on one-forms, but it does not require an extra cotangent
direction as in Ref. \cite{Sitarz}. When the action is restricted
to $\kappa$-Minkowski space (which is a subalgebra od
$\mathfrak{D}_1$ and $\mathfrak{D}_2$), then the Lorentz algebra
acts covariantly on products of space-time coordinates which
reproduces the well-know result of Majid and
Ruegg \cite{Majid-Ruegg}. In Sec. 4 a short conclusion and future
outlook is given.

\section{Differential algebras on $\kappa$-Minkowski space}

In this section we present the main points of the construction of
differential algebras on $\kappa$-Minkowski space using
realizations. First, we consider the differential algebra
introduced by Sitarz \cite{Sitarz} (see also Ref. \cite{Majid-2}).

The $\kappa$-Minkowski space $\mathcal{M}_\kappa$ is an associative algebra generated by space-time coordinates $\hat x_0,
\hat x_1, \ldots , \hat x_{n-1}$ satisfying the Lie algebra type commutation relations
\begin{equation}\label{1}
[\hat x_i,\hat x_j]=0, \quad [\hat x_0,\hat x_j]=ia_0\, \hat x_j, \quad a_0\in \R.
\end{equation}
By convention latin indices run from $1$ to $n-1$, and greek
indices run from  $0$ to $n-1$. A bicovariant differential algebra
compatible with relations \eqref{1} was constructed by Sitarz in
Ref. \cite{Sitarz}. He has shown that if the differential algebra
is also Lorentz covariant, then the smallest such algebra in 3+1
dimensions is five-dimensional. One of its equivalent forms is
given by
\begin{alignat}{3}
[\hat \xi_0,\hat x_0] &= -ia_0\, \theta^\prime+ia_0\, \hat \xi_0, \quad & [\hat \xi_0,\hat x_j] &= ia_0\, \hat \xi_j, \label{2} \\
[\hat \xi_i,\hat x_0] &= 0, \quad & [\hat \xi_i,\hat x_j] &= -ia_0\, \delta_{ij}\, \theta^\prime, \\
[\theta^\prime, \hat x_0] &= -ia_0\, \theta^\prime, \quad & [\theta^\prime, \hat x_j] &=0, \label{4}
\end{alignat}
where $\hat \xi_\mu$ is the one-form corresponding to $\hat
x_\mu$, and $\theta^\prime$ is a one-form representing an extra
cotangent direction that has no classical analogue. The one-forms
$\hat\xi_\mu$ and $\theta^\prime$ anticommute. Let $\mathfrak{D}$
denote the algebra \eqref{2}-\eqref{4}. The algebra $\mathfrak{D}$
was considered in Ref. \cite{Majid-2} where it is shown that by
gauging a coefficient of $\theta^\prime$ one  can introduce
gravity in the model. If we make a change of basis $\theta=\hat
\xi_0-\theta^\prime$, we recover the original algebra introduced
in Ref. \cite{Sitarz}. As stated above, $\mathfrak{D}$ is
constructed by postulating both bicovariance and Lorentz
covariance of the differential calculus on $\kappa$-Minkowski
space. These conditions imply that
\begin{enumerate}[(1)]
\item $[\hat x_\mu, \hat \xi_\nu]$ and $[\hat x_\mu, \theta^\prime]$ are closed in the vector space spanned by one-forms alone,
\item all graded Jacobi identities in $\mathfrak{D}$ hold,
\item the action of the Lorentz algebra $\mathfrak{so}(1,n-1)$ is covariant:
\end{enumerate}
\begin{align}
M\blacktriangleright (\hat x_\mu \hat \xi_\nu) &= \big(M_{(1)}\blacktriangleright \hat x_\mu\big)\, \hat d\big(M_{(2)}
\blacktriangleright \hat x_\nu\big), \label{5} \\
M\blacktriangleright (\hat \xi_\mu \hat x_\nu) &= \hat
d\big(M_{(1)}\blacktriangleright \hat x_\mu\big)
\big(M_{(2)}\blacktriangleright \hat x_\nu\big), \quad
M\blacktriangleright \theta^\prime =0,
\end{align}
where $\hat d$ is the exterior derivative and $M$ is a generator of $\mathfrak{so}(1,n-1)$. Here, the commutation relations in $\mathfrak{so}(1,n-1)$
are undeformed,
\begin{equation}\label{7}
[M_{\mu\nu},M_{\lambda\rho}]=\eta_{\nu\lambda}M_{\mu\rho}-\eta_{\mu\lambda}M_{\nu\rho}-\eta_{\nu\rho}M_{\mu\lambda}+\eta_{\mu\rho}M_{\nu\lambda},
\end{equation}
and $\Delta M=M_{(1)}\otimes M_{(2)}$ is the coproduct of $M$ in Sweedler notation:
\begin{align}
\Delta M_{i0} &= M_{i0} \otimes 1 + e^{a_0 p_0} \otimes M_{i0}-a_0 \sum_{j=1}^{n-1} p_j \otimes M_{ij}, \label{8} \\
\Delta M_{ij} &= M_{ij} \otimes 1 + 1\otimes M_{ij},  \label{9}
\end{align}
where $p_\mu$ is the momentum generator. The coproduct of the momentum generators is given by \cite{Sitarz}, \cite{Meljanac-2}
\begin{align}
\Delta p_0 &= p_0 \otimes 1 + 1\otimes p_0, \label{10A} \\
\Delta p_i &= p_i \otimes 1 + e^{a_0 p_0}\otimes p_i. \label{11A}
\end{align}
Relations \eqref{8}-\eqref{11A} describe the coalgebra structure of the $\kappa$-Poincar\'{e} algebra generated by $M_{\mu\nu}$ and $p_\mu$.
 Note that Eq. \eqref{5} implies that the Lorentz generators act on a constant one-form by
$M\blacktriangleright \hat \xi_\mu = \hat
d\big(M\blacktriangleright \hat x_\mu\big)$.

In the following we shall briefly outline the construction of the algebra $\mathfrak{D}$ using realizations of $\hat x_\mu$, $\hat \xi_\mu$ and
$\theta^\prime$ as formal power series in a Weyl super-algebra. Let $\mathcal{A}$ denote the unital associative
algebra generated by commutative coordinates $x_\mu$, differential operators $\p_\mu=\frac{\p}{\p x_\mu}$ and ordinary one-forms
$dx_\mu$ satisfying the defining relations
\begin{alignat}{2}
[x_\mu,x_\nu] &= [\p_\mu,\p_\nu]=0, \qquad & [\p_\mu,x_\nu] &= \eta_{\mu\nu}, \label{10}\\
[dx_\mu, x_\nu] &= [dx_\mu,\p_\nu] = 0, \qquad & \{dx_\mu, dx_\nu\} &= 0.  \label{11}
\end{alignat}
Here, $\{\, ,\, \}$ denotes the anticommutator and $\eta=diag(-1,1,\ldots, 1)$ is the Minkowski metric. $\mathcal{A}$ becomes a
Weyl super-algebra if we define a graded commutator
\begin{equation}
[[u,v]]=uv-(-1)^{|u|\, |v|} vu
\end{equation}
where $|u|$ denotes the degree of a homogeneous element $u\in
\mathcal{A}$. The degrees of the generators are defined by
$|x_\mu|=|\p_\mu|=0$ and $|dx_\mu|=1$. In this paper we consider
two types of realizations of $\hat x_\mu$, the natural \cite{Meljanac-3} and
noncovariant \cite{Meljanac-4}, \cite{ms06}. Following the
notation in Ref. \cite{Meljanac-2} the variables used in the
natural and noncovariant realizations are denoted by
$(X_\mu,D_\mu)$ and $(x_\mu,\p_\mu)$, respectively. The reason for
using different notation for the generators of $\mathcal{A}$ is
that there exists an invertible transformation
$(x_\mu,\p_\mu)\mapsto (X_\mu,D_\mu)$ mapping the noncovariant
into natural realization. The natural realization is defined by
\begin{equation}\label{12}
\hat x_\mu = X_\mu Z^{-1}-ia_0 X_0 D_\mu
\end{equation}
where $Z$ is invertible operator given by
\begin{equation}\label{14}
Z^{-1} = ia_0\, D_0 + \sqrt{1+a_0^2\, D^2}.
\end{equation}
The scalar product in \eqref{14} is taken with respect to the  Minkowski metric, i.e. $D^2 = -D_0^2+\sum_{k=1}^{n-1} D_k^2$.
$Z$ is called the shift operator because conjugation of $\hat x_\mu$ by $Z$ yields $Z \hat x_\mu Z^{-1} = \hat x_\mu + ia_0 \delta_{0\mu}$.
One easily checks that the space-time coordinates represented by \eqref{12} satisfy the commutation relations \eqref{1}. The realization \eqref{12}
is a special case of covariant realizations of $\kappa$-Minkowski space introduced in Ref. \cite{Meljanac-3}.\\

Exterior derivative $\hat d$ is defined by
$\hat d = \sum_{\alpha, \beta=0}^{n-1} k_{\alpha \beta}(D)\, dX_\alpha\, D_\beta$ where $k_{\alpha\beta}(D)$ is a formal
power series in $a_0$ with coefficients in the ring of differential operators $D_\mu$. We require that $\lim_{a_0\to 0} \hat d= d$ where
$d=-dX_0\, D_0 + \sum_{k=1}^{n-1} dX_k\, D_k$ is the classical exterior derivative.
The exterior derivative acts on space-time coordinates by $\hat d \cdot \hat x_\mu = [[\hat d,\hat x_\mu]]$. We define a noncommutative
version of one-forms by $\hat \xi_\mu = \hat d \cdot \hat x_\mu$. Using relations \eqref{10}-\eqref{11} we find
\begin{equation}
\lim_{a_0\to 0} \hat \xi_\mu = [d,X_\mu]=dX_\mu,
\end{equation}
hence $\hat \xi_\mu$ is a deformation of ordinary one-form $dX_\mu$. Before proceeding further let us point out some
general properties of a differential algebra constructed in this way:
\begin{enumerate}[(1)]
\item $\hat d$ satisfies the undeformed Leibniz rule
\begin{equation}
\hat d\cdot (f(\hat x) g(\hat x)) = (\hat d\cdot f(\hat x)) g(\hat x) + f(\hat x) (\hat d \cdot g(\hat x))
\end{equation}
where $f(\hat x)$ and $g(\hat x)$ are monomials in $\hat x_\mu$.
\item one-forms are closed, i.e. $\hat d \cdot \hat \xi_\mu = [[\hat d, \hat \xi_\mu]]=0$,
\item one-forms anticommute, $\{\hat \xi_\mu,\hat \xi_\nu\}=0$,
\item the commutator for $\hat \xi_\mu$ and $\hat x_\nu$ is given by
\begin{equation}
[\hat \xi_\mu,\hat x_\nu] = \sum_{\alpha=0}^{n-1} K^\alpha_{\mu\nu}(D)\, \hat \xi_\alpha
\end{equation}
where $K^\alpha_{\mu\nu}(D)$ generally depends on the differential operators $D_\mu$. If $K^\alpha_{\mu\nu}$ are constant for all values of
$\mu,\nu$ and $\alpha$, then the differential algebra is closed.
\end{enumerate}
 We note that the Jacobi identity for $\hat d\cdot [\hat x_\mu,\hat x_\nu]=[\hat d,[\hat x_\mu,\hat x_\nu]]$ together with commutation
relations \eqref{1} implies that $\hat x_\mu$ and $\hat \xi_\mu$ satisfy the compatibility condition
\begin{equation}\label{17AA}
[\hat \xi_\mu, \hat x_\nu]-[\hat \xi_\nu, \hat x_\mu]=i(a_\mu \hat x_\nu-a_\nu \hat \xi_\mu)
\end{equation}
where $a_\mu = a_0 \delta_{0\mu}$. Extension of the above
construction to higher order forms was presented in detail in Ref.
\cite{Meljanac-1}.\

Given the realization \eqref{12} we want to find a realization of $\hat d$ such that the action of $\hat d$
on $\hat x_\mu$ generates one-forms $\hat \xi_\mu$ and $\theta^\prime$ which close the algebra \eqref{2}-\eqref{4}.
Consider the following ansatz for $\hat d$:
\begin{equation}\label{17A}
\hat d = -dX_0\, D_0 + \big(\sum_{k=1}^{n-1} dX_k\, D_k\big) Z.
\end{equation}
Substituting Eqs. \eqref{12} and \eqref{17A} into $\hat \xi_\mu = [[\hat d, \hat x_\mu]]$ one finds
\begin{align}
\hat \xi_0 &= dX_0 \big(Z^{-1}-ia_0 D_0\big)+ia_0 \big(\sum_{k=1}^{n-1} dX_k\, D_k \big) Z, \\
\hat \xi_k &= dX_k - ia_0\, dX_0\, D_k.
\end{align}
The commutation relations for $\hat x_\mu$ and $\hat \xi_\mu$ are given by
\begin{alignat}{2}
[\hat \xi_0, \hat x_0] &= -ia_0\, dX_0\, Z^{-1}+ia_0 \hat \xi_0, \quad &  [\hat \xi_0,\hat x_j] &= ia_0\, \hat \xi_j,  \label{20A} \\
[\hat \xi_i, \hat x_0] &= 0, \quad & [\hat \xi_i,\hat x_j] &= -ia_0\, \delta_{ij}\, dX_0\, Z^{-1}.  \label{21A}
\end{alignat}
Note that the algebra \eqref{20A}-\eqref{21A} is not closed since the commutators involve an additional term $dX_0\, Z^{-1}$ which does not
correspond to any one-form $\hat \xi_\mu$. However, the algebra can be closed by defining an extra one-form by
$\theta^\prime = dX_0\, Z^{-1}$. Then one easily finds
\begin{equation}
[\theta^\prime, \hat x_0] = -ia_0 \theta^\prime, \quad [\theta^\prime, \hat x_j]=0. \label{22A}
\end{equation}
The commutation relations \eqref{20A}-\eqref{22A} agree with the
differential algebra \eqref{2}-\eqref{4}. Thus, in our approach
the extra cotangent direction $\theta^\prime$ introduced in Refs.
\cite{Sitarz} and \cite{Majid-2} appears as a deformation of
one-form $dX_0$ associated with time coordinate. In fact, $\hat
\xi_0$ and $\theta^\prime$ are both deformations of $dX_0$, albeit
different.

\subsection{Differential algebras of classical dimension}

Different realizations of $\hat x_\mu$ and $\hat d$ lead to different differential calculi on $\kappa$-Minkowski space $\mathcal{M}_\kappa$.
In the following we construct two families of differential algebras on $\mathcal{M}_\kappa$ such that there is a
one-to-one correspondence between deformed one-forms and space-time coordinates. In both cases the realization of $\hat x_\mu$ is the same,
but we consider two different deformations of the exterior derivative $\hat d$.

Let
\begin{equation}\label{24A}
\hat x_0 = x_0 + ia_0 \sum_{k=1}^{n-1} x_k \p_k, \quad \hat x_k = x_k.
\end{equation}
This is a special case of noncovariant realizations of the algebra
\eqref{1} introduced in Ref. \cite{Meljanac-4}, \cite{ms06}. The
transformation of variables $(x_\mu,\p_\mu)\mapsto (X_\mu,D_\mu)$
which maps the noncovariant into natural realization is given in
Ref. \cite{Meljanac-2}. The realization \eqref{24A} corresponds to
the bicrossproduct basis in Ref. \cite{Majid-Ruegg}. The shift
operator corresponding to realization \eqref{24A} is given by
\begin{equation}\label{25}
Z=\exp(A), \quad A=-ia_0 \p_0.
\end{equation}
Let us define the exterior derivative
\begin{equation}
\hat d_1 = dx_0 \frac{Z^{c}-1}{ia_0\, c}+\Big(\sum_{k=1}^{n-1} dx_k \p_k\Big) Z^{-1}, \quad c\neq 0,
\end{equation}
where for $c\to 0$ we have $\lim_{c\to 0}\hat{d}_{1}=-dx_{0}\partial_{0}+\big(\sum_{k=1}^{n-1} dx_k \p_k\big) Z^{-1}$.
 The one-forms $\hat \xi_\mu = [[\hat d_1,\hat x_\mu]]$ with space-time coordinates represented by \eqref{24A} are
given by
\begin{equation}\label{27AA}
\hat \xi_0 = dx_0 Z^c, \quad \hat \xi_k = dx_k Z^{-1}.
\end{equation}
Using realizations \eqref{24A} and \eqref{27AA} we find
\begin{equation}
[\hat \xi_0,\hat x_0] = ia_0\, c\, \hat \xi_0, \quad  [\hat \xi_k, \hat x_0] = -ia_0\, \hat \xi_k, \quad [\hat \xi_\mu,\hat x_j]=0, \label{28AA} \\
\end{equation}
We denote the differential algebra \eqref{28AA} by $\mathfrak{D}_1$. Similarly,
if the exterior derivative is defined by
\begin{equation}
\hat d_2 = dx_0 \frac{Z^c-1}{ia_0\, c}+\Big(\sum_{k=1}^{n-1} dx_k\, \p_k\Big) Z^{c-1},
\end{equation}
then the one-forms $\hat \xi_\mu=[[\hat d_2,\hat x_\mu]]$ are found to be
\begin{equation}\label{31A}
\hat \xi_0 = dx_0 Z^c + ia_0\, c\big(\sum_{k=1}^{n-1} dx_k \p_k\big) Z^{c-1}, \quad \hat \xi_k = dx_k Z^{c-1}.
\end{equation}
Now Eqs. \eqref{24A} and \eqref{31A} imply
\begin{equation}
[\hat \xi_0, \hat x_\mu] = ia_0\, c\,\hat \xi_\mu, \quad
[\hat \xi_k, \hat x_0] = ia_0 (c-1) \hat \xi_k, \quad
[\hat \xi_k, \hat x_j] = 0.  \label{34A}
\end{equation}
The differential algebra \eqref{34A} is denoted by $\mathfrak{D}_2$. Note that $\mathfrak{D}_1$ and $\mathfrak{D}_2$ are two families of
differential algebras depending on a real parameter $c$ obtained from a fixed realization of $\hat x_\mu$. They
are compatible with $\kappa$-Minkowski space since they satisfy the compatibility condition \eqref{17AA} and all graded
Jacobi identities for the generators of $\mathfrak{D}_1$ and $\mathfrak{D}_2$ hold. The commutator $[\hat \xi_\mu, \hat x_\nu]$
in both algebras is closed in the vector space spanned by $\hat \xi_0, \hat \xi_1, \ldots ,\hat \xi_{n-1}$, hence
$\mathfrak{D}_1$ and $\mathfrak{D}_2$ are differential algebras of classical dimension.

\section{Action of the Lorentz algebra on $\mathfrak{D}_1$ and $\mathfrak{D}_2$}

The aim of this section is to construct an action of the Lorentz algebra $\mathfrak{so}(1,n-1)$ on the algebras $\mathfrak{D}_1$ and
$\mathfrak{D}_2$. First, we define an action of $\mathfrak{so}(1,n-1)$ on the subalgebra $\mathcal{M}_\kappa$ and then extend it to
$\mathfrak{D}_1$ and $\mathfrak{D}_2$.\\

It is natural to consider extension of the $\kappa$-Minkowski space \eqref{1} by momentum operators $p_\mu$. If we take the realization
\eqref{24A} and define $p_\mu = -i\p_\mu$, then $\hat x_\mu$ and $\p_\mu$ generate a deformed Heisenberg algebra given by the
relations \eqref{1} and
\begin{equation}\label{33}
[p_\mu,p_\nu]=0, \quad [p_0,\hat x_\mu]=i\delta_{o\mu}, \quad [p_k,\hat x_0]=ia_0 p_k, \quad [p_k,\hat x_j]=-i\delta_{kj}.
\end{equation}
Note that the deformation of the algebra \eqref{33} depends on the
realization of the Minkowski coordinates $\hat x_\mu$. A large
class of such deformations was found in Refs. \cite{Meljanac-3},
\cite{Meljanac-4} and \cite{ms06}. Similarly, the
$\kappa$-Minkowski space can be extended by the Lorentz algebra
such that the direct sum of vector spaces
$\mathfrak{g}_\kappa=\mathcal{M}_\kappa\oplus
\mathfrak{so}(1,n-1)$ is a Lie algebra. It can be shown that the
cross commutator $[M_{\mu\nu},\hat x_\lambda]$, which must be
linear in $M_{\mu\nu}$ and $\hat x_\mu$, is uniquely given
by \cite{Meljanac-4}, \cite{ms06}
\begin{alignat}{2}
[M_{i0},\hat x_0] &= -\hat x_i + ia_0 M_{i0}, \qquad & [M_{ij}, \hat x_0] &= 0, \label{35} \\
[M_{i0},\hat x_k] &= -\delta_{ik} \hat x_0+ia_0 M_{ik}, \qquad & [M_{ij}, \hat x_k] &= \delta_{jk} \hat x_i-\delta_{ik} \hat x_j. \label{36}
\end{alignat}
The algebra \eqref{35}-\eqref{36} is a subalgebra of the DSR algebra obtained as a cross product extension of $\kappa$-Minkowski and
$\kappa$-Poincar\'{e} algebras \cite{Borowiec-1}, \cite{Borowiec-2}. Since the commutation relations \eqref{35}-\eqref{36}
are unique, the extension of $\mathcal{M}_\kappa$ by $\mathfrak{so}(1,n-1)$ is independent of the realization of $\hat x_\mu$.
If the coordinates $\hat x_\mu$ are given by the noncovariant realization \eqref{24A}, then the Lorentz
generators are represented by
\begin{align}
M_{i0} &= x_i \left(\frac{1-Z}{ia_0}+\frac{ia_0}{2}\Delta - \frac{2}{ia_0} \sinh^2\big(\frac{1}{2}A\big) Z\right)
-\Big(x_0+ia_0 \sum_{k=1}^{n-1} x_k \p_k\Big) \p_i, \label{37}  \\
M_{ij} &= x_i \p_j - x_j \p_i, \label{38}
\end{align}
where $Z$ is given by Eq. \eqref{25} and $\Delta =\sum_{k=1}^{n-1}
\p_k^2$ is the Laplace operator. The realization
\eqref{37}-\eqref{38} is a special case of the noncovariant
realizations of the Lorentz algebra found in Refs.
\cite{Meljanac-4}, \cite{ms06}. In the classical limit we have
$\lim_{a_0\to 0} M_{\mu\nu} = x_\mu\p_\nu - x_\nu \p_\mu$, as
required.

Given the commutation relations \eqref{1} and
\eqref{35}-\eqref{36} we want to define an action
$\blacktriangleright \colon \mathfrak{so}(1,n-1)\times
\mathcal{M}_\kappa \to \mathcal{M}_\kappa$. Let
$U(\mathfrak{g}_\kappa)$ be the enveloping algebra of
$\mathfrak{g}_\kappa$ and let the generators of
$U(\mathfrak{g}_\kappa)$ act on $1\in U(\mathfrak{g}_\kappa)$ by
$\hat x_\mu \blacktriangleright 1 = \hat x_\mu$ and
$M_{\mu\nu}\blacktriangleright 1 = 0$. Now define
\begin{equation}\label{39}
M_{\mu\nu}\blacktriangleright f(\hat x) = [M_{\mu\nu},f(\hat
x)]\blacktriangleright 1
\end{equation}
where $f(\hat x)$ is a monomial in $\mathcal{M}_\kappa$. Using relations \eqref{35}-\eqref{36} the commutator in \eqref{39} can be
written as a linear combination of terms with $M_{\mu\nu}$, if any, pushed to the far right. Thus, the action \eqref{39} is the projection of
$[M_{\mu\nu},f(\hat x)]$ to the subalgebra $\mathcal{M}_\kappa$. For example, the action on Minkowski coordinates yields
\begin{equation}\label{38A}
M_{\mu\nu}\blacktriangleright \hat x_\lambda = \eta_{\nu\lambda}
\hat x_\mu - \eta_{\mu\lambda} \hat x_\nu.
\end{equation}
For monomials of order two we find
\begin{align}
M_{i0} \blacktriangleright (\hat x_0\hat x_k) &= -\hat x_i \hat x_k -ia_0 \delta_{ik} \hat x_0 - \delta_{ik} \hat x_0^2,  \\
M_{i0} \blacktriangleright (\hat x_k \hat x_0) &= -\hat x_k \hat x_i-\delta_{ik} \hat x_0^2, \\
M_{i0} \blacktriangleright (\hat x_k \hat x_l) &= \delta_{ik} \hat x_0 \hat x_l -\delta_{il} \hat x_k \hat x_0
+ia_0(\delta_{kl}\hat x_i-\delta_{il}\hat x_k), \\
M_{ij} \blacktriangleright (\hat x_0 \hat x_k) &= \delta_{jk} \hat x_0 \hat x_i - \delta_{ik} \hat x_0\hat x_j, \\
M_{ij} \blacktriangleright (\hat x_k \hat x_0) &= \delta_{jk} \hat x_i \hat x_0 - \delta_{ik} \hat x_j \hat x_0, \\
M_{ij} \blacktriangleright (\hat x_k \hat x_l) &= \delta_{jk} \hat
x_i \hat x_l -\delta_{ik} \hat x_j \hat x_l + \delta_{jk} \hat x_k
\hat x_i- \delta_{il} \hat x_k \hat x_j.
\end{align}
The above result is the same as that obtained by Majid and
Ruegg \cite{Majid-Ruegg} using the covariance condition $M_{\mu\nu}
\blacktriangleright (ab) = (M_{{\mu\nu}_{(1)}}\blacktriangleright
a) (M_{{\mu\nu}_{(2)}}\blacktriangleright b)$, $a,b\in
\mathcal{M}_\kappa$, where the coproduct $\Delta M_{\mu\nu} =
M_{{\mu\nu}_{(1)}} \otimes M_{{\mu\nu}_{(2)}}$ is given by Eqs.
\eqref{8}-\eqref{9} and the momentum operator in \eqref{8} acts by
$p_\mu\blacktriangleright \hat x_\nu = -i\eta_{\mu\nu}$. This
makes $\mathcal{M}_\kappa$ into an $\mathfrak{so}(1,n-1)$-module
algebra.

Next, we want to extend the action of $\mathfrak{so}(1,n-1)$ to the differential algebras $\mathfrak{D}_1$ and $\mathfrak{D}_2$ using the same
prescription \eqref{39}. For this purpose we need the commutator $[M_{\mu\nu},\hat \xi_\lambda]$ where $\hat \xi_\lambda\in \mathfrak{D}_1$
or $\hat \xi_\lambda\in \mathfrak{D}_2$. One can show that the general form of this commutator is given by
\begin{equation}\label{48}
[M_{\mu\nu},\hat \xi_\lambda] = [\hat x_\mu,\hat \xi_\lambda] \Phi_\nu - [\hat x_\nu, \hat \xi_\lambda]\Phi_\mu
\end{equation}
where $\Phi_\mu$ is a power series in $\p_\mu$ such that $\Phi(0)=0$. The functions $\Phi_\mu$ depend on the realization of
$M_{\mu\nu}$. If $M_{\mu\nu}$ is given by Eqs. \eqref{37}-\eqref{38}, then
\begin{align}
\Phi_0 &= \frac{1-Z}{ia_0}+\frac{ia_0}{2}\Delta - \frac{2}{ia_0} \sinh^2\big(\frac{1}{2}A\big) Z, \\
\Phi_k &= \p_k,
\end{align}
where $\Delta = \sum_{k=1}^{n-1} \p^2_k$. Since $[\hat x_\mu,\hat \xi_\nu]\in span\{\hat \xi_\mu\mid 0\leq \mu \leq n-1\}$, the
commutator $[M_{\mu\nu},\hat \xi_\lambda]$ depends only on momentum operators $p_\mu = -i\p_\mu$ and one-forms $\hat
\xi_\lambda$. This means that in order to extend the action of $\mathfrak{so}(1,n-1)$ to $\mathfrak{D}_k$ we need to extend the
algebra $U(\mathfrak{g}_\kappa)$ by the generators $p_\mu$ and $\hat \xi_\mu$ where the extension depends on whether $\hat
\xi_\mu \in\mathfrak{D}_1$ or $\hat \xi_\mu\in \mathfrak{D}_2$. Denote the extended algebras by $\mathfrak{H}_1$ and
$\mathfrak{H}_2$, respectively. Then $\mathfrak{H}_k$ contains $\mathfrak{D}_k$ and $\mathfrak{so}(1,n-1)$ as Lie subalgebras, as
well as the abelian algebra of translations generated by $p_\mu$. Define the action of $p_\mu$ and $\hat \xi_\mu$ on $1\in
\mathfrak{H}_k$ by $p_\mu \blacktriangleright 1=0$ and $\hat \xi_\mu\blacktriangleright 1 =\hat \xi_\mu$. Now we may define the
action $\blacktriangleright \colon \mathfrak{so}(1,n-1)\times \mathfrak{D}_k \to \mathfrak{D}_k$ by
\begin{equation}\label{51}
M_{\mu\nu}\blacktriangleright f(\hat x,\hat
\xi)=[M_{\mu\nu},f(\hat x,\hat \xi)]\blacktriangleright 1
\end{equation}
where $f(\hat x,\hat \xi)$ is a monomial in $\mathfrak{D}_k$, $k=1,2$. The action \eqref{51} is uniquely fixed by the
commutation relations in $\mathfrak{D}_k$ and Eqs. \eqref{35}-\eqref{36} and \eqref{48}. Since the generators of
$\mathfrak{H}_k$ are constructed as elements of an associative algebra all Jacobi identities in $\mathfrak{H}_k$ hold. The Jacobi
relations for $M_{\mu\nu}$, $\hat x_\mu$ and $\hat \xi_\mu$ guarantee that the action \eqref{51} is compatible with the
commutation relations in $\mathfrak{D}_k$. Any monomial in $\mathfrak{D}_k$ can be written as a finite sum $f(\hat x,\hat
\xi)=\sum f_1(\hat x) f_2(\hat \xi)$, hence it suffices to consider the action \eqref{51} on the products $f_1(\hat x)
f_2(\hat \xi)$. Since $\Phi_{\mu}\blacktriangleright 1=0$, Eq. \eqref{48} implies that the action of $M_{\mu\nu}$ on one-forms is
trivial, i.e. $M_{\mu\nu}\blacktriangleright f(\hat \xi)=0$ for any monomial $f(\hat \xi)$. Consequently, the action of
$M_{\mu\nu}$ on product of monomials $f_1(\hat x) f_2(\hat \xi)$ is given by
\begin{equation}
M_{\mu\nu}\blacktriangleright f_1(\hat x) f_2(\hat \xi) =
\big(M_{\mu\nu} \blacktriangleright f_1(\hat x)\big) f_2(\hat \xi).
\end{equation}
The construction outlined here has the advantage that the action
\eqref{51} is compatible with the algebra structure of
$\mathfrak{D}_1$ and $\mathfrak{D}_2$ without introducing the
extra one-form $\theta$ as in Ref. \cite{Sitarz}. However, the
action is not Lorentz covariant since the necessary condition
$M_{\mu\nu}\blacktriangleright \hat \xi_\lambda =\hat d\cdot
(M_{\mu\nu}\blacktriangleright \hat x_\lambda)$ does not hold.
Indeed, in view of Eq. \eqref{38A} we have $\hat d \cdot
(M_{\mu\nu} \blacktriangleright \hat x_\lambda)=\eta_{\nu\lambda}
\hat \xi_\mu - \eta_{\mu\lambda} \hat\xi_\nu$ while
$M_{\mu\nu}\blacktriangleright \hat \xi_\lambda =0$. This problem
can be partially resolved by modifying the realization of
$M_{\mu\nu}$ such that $M_{\mu\nu}\blacktriangleright \hat
\xi_\lambda = \eta_{\nu\lambda}\hat \xi_\mu-\eta_{\mu\lambda} \hat
\xi_\nu$ is satisfied. This modification of \eqref{51} if  given
as follows.

Consider extension $\tilde{\mathcal{A}}$ of the algebra \eqref{10}-\eqref{11} by a set of generators $q_\mu$ subject to defining relations
\begin{equation}
[x_\mu,q_\nu]=[\p_\mu,q_\nu]=0, \quad \{q_\mu, q_\nu\}=0, \quad \{dx_\mu,q_\nu\}=\eta_{\mu\nu}.
\end{equation}
The degree of $q_\mu$ is defined to be $|q_\mu|=1$. Note that the variables $q_\mu$ play the role of a Grassmann type derivative with
respect to one-forms $dx_\mu$. The $\kappa$-deformed super-Heisenberg algebra, generated by $\hat{x}_{\mu},\, \partial_{\mu},\, \hat{\xi}_{\mu}$
and $q_{\mu}$, satisfies all graded Jacobi identities. Let us define
\begin{equation}\label{551}
M_{\mu\nu}^{(1)}= dx_\mu\, q_\nu - dx_\nu\, q_\mu,
\end{equation}
and let
\begin{equation}\label{52}
\widetilde M_{\mu\nu} = M_{\mu\nu} + M_{\mu\nu}^{(1)}
\end{equation}
where the Lorentz generators $M_{\mu\nu}$ are given by the realization \eqref{37}-\eqref{38}. It is easily seen that $M_{\mu\nu}^{(1)}$
close the relations \eqref{7} and $[M_{\mu\nu}, M_{\lambda\rho}^{(1)}]=0$. Consequently, $\widetilde M_{\mu\nu}$ also satisfy the
relations \eqref{7}, hence Eq. \eqref{52} represents a new realization of $\mathfrak{so}(1,n-1)$ in terms of the extended algebra
$\tilde{\mathcal{A}}$. To this realization of $\mathfrak{so}(1,n-1)$ we associate the action
\begin{equation}\label{53}
\widetilde M_{\mu\nu}\blacktriangleright f(\hat x,\hat \xi) =
[\widetilde M_{\mu\nu},f(\hat x, \hat \xi)]\blacktriangleright 1.
\end{equation}
The action is consistent with the commutation relations in $\mathfrak{D}_1$ and $\mathfrak{D}_2$ since all Jacobi identities
for $\widetilde M_{\mu\nu}$, $\hat x_\mu$ and $\hat \xi_\mu$ are satisfied.
For products of monomials $f_1(\hat x) f_2(\hat \xi)$ the action \eqref{53} satisfies the Leibniz-like rule
\begin{equation}\label{54}
\widetilde M_{\mu\nu}\blacktriangleright f_1(\hat x) f_2(\hat
\xi)=\big(M_{\mu\nu}\blacktriangleright f_1(\hat x)\big)f_2(\hat
\xi)+f_1(\hat x)\big(M_{\mu\nu}^{(1)} \blacktriangleright f_2(\hat
\xi)\big)
\end{equation}
where $M_{\mu\nu}$ acts only on coordinates $\hat x_\mu$ and $M_{\mu\nu}^{(1)}$ acts only on one-forms $\hat \xi_\mu$. It
follows from Eq. \eqref{54} that $\widetilde M_{\mu\nu}\blacktriangleright f_1(\hat x) = M_{\mu\nu}
\blacktriangleright f_1(\hat x)$, hence the actions \eqref{51} and \eqref{53} agree on the $\kappa$-Minkowski space. In particular,
we have the vector-like transformation $\widetilde M_{\mu\nu} \blacktriangleright \hat x_\lambda = \eta_{\nu\lambda} \hat x_\mu
-\eta_{\mu\lambda} \hat x_\nu$. On the other hand, the action of $\widetilde M_{\mu\nu}$ on one-forms is nontrivial since one-forms
also transform vector-like, $\widetilde M_{\mu\nu}\blacktriangleright \hat \xi_\lambda = \eta_{\nu\lambda}
\hat \xi_\mu-\eta_{\mu\lambda}\hat \xi_\nu$. This implies
\begin{equation}\label{55}
\widetilde M_{\mu\nu} \blacktriangleright \hat\xi_\lambda =\hat d
\cdot (\widetilde M_{\mu\nu} \blacktriangleright \hat x_\lambda),
\end{equation}
thus the action is Lorentz covariant on constant one-forms. Restriction of \eqref{53} to monomials in
$\hat \xi_\mu$ satisfies the ordinary Leibniz rule
\begin{equation}
\widetilde M_{\mu\nu} \blacktriangleright f(\hat \xi) g(\hat
\xi)=\big(\widetilde M_{\mu\nu}\blacktriangleright f(\hat\xi)\big)
g(\hat \xi)+ f(\hat \xi)\big(\widetilde
M_{\mu\nu}\blacktriangleright g(\hat \xi)\big).
\end{equation}
Using Eq. \eqref{54} and the rules for computing $M_{\mu\nu}\blacktriangleright f_1(\hat x)$ and
$M_{\mu\nu}^{(1)}\blacktriangleright f_2(\hat \xi)$ one can easily calculate the action of $\widetilde M_{\mu\nu}$ on arbitrary
monomials $f(\hat x,\hat \xi)\in \mathfrak{D}_k$, $k=1,2$. For example,
\begin{equation}\label{57}
\widetilde M_{\mu\nu}\blacktriangleright \hat x_\lambda \hat
\xi_\rho = (\eta_{\nu\lambda}\hat x_\mu-\eta_{\mu\lambda}\hat
x_\nu) \hat\xi_\rho+ \hat x_\lambda (\eta_{\nu\rho} \hat
\xi_\mu-\eta_{\mu\rho} \hat\xi_\nu).
\end{equation}
% Odlomak poslije ove jednadzbe je izbrisan jer je nepotreban.

The condition \eqref{55} does not extend by Lorentz covariance to
entire algebras $\mathfrak{D}_1$ and $\mathfrak{D}_2$. This is in
accordance with the theory developed in Ref. \cite{Sitarz} since
otherwise this would be in contradiction with the Jacobi identity
for $\hat x_\mu$, $\hat x_\nu$ and $\hat \xi_\lambda$ (for more
details see Ref. \cite{Sitarz}).

\section{Conclusion}

In this paper we have constructed differential algebras on
$\kappa$-Minkowski space-time using realizations of coordinates
$\hat x_\mu$ and one-forms $\hat \xi_\mu$ as formal power series
in a Weyl super-algebra. The algebras considered here are the
well-known differential algebra introduced by Sitarz \cite{Sitarz}
as well as new families of differential algebras $\mathfrak{D}_1$
and $\mathfrak{D}_2$. The algebras $\mathfrak{D}_1$ and
$\mathfrak{D}_2$ are obtained from a fixed realization of $\hat
x_\mu$ and using different realization of exterior derivative
$\hat d$. The resulting one-forms $\hat \xi_\mu=[[\hat d,\hat
x_\mu]]$ have the property that the commutator $[\hat \xi_\mu,
\hat x_\nu]$ is closed in the vector space spanned by $\hat \xi_0,
\hat \xi_1,\ldots ,\hat \xi_{n-1}$ alone. We have also presented a
novel construction of an action of $\mathfrak{so}(1,n-1)$ on
$\mathfrak{D}_1$ and $\mathfrak{D}_2$ using realizations of the
Lorentz generators in terms of Grassmann-type variables $q_\mu$.
The action does not require introduction of an extra cotangent
direction $\theta$ as in Ref. \cite{Sitarz}. When restricted to
Minkowski coordinates, $\mathfrak{so}(1,n-1)$ acts covariantly on
the $\kappa$-Minkowski space making it into an
$\mathfrak{so}(1,n-1)$-module algebra. The Lorentz covariance is
valid  for constant one-forms but it does not extend to entire
algebras $\mathfrak{D}_1$ and $\mathfrak{D}_2$. In this approach
there is a one-to-one correspondence between the Minkowski
coordinates and one-forms. This provides a certain advantage since
every variable in the noncommutative setting is for a given
realization a unique deformation of the corresponding classical
variable. In this paper we have focused only on the action of the
Lorentz algebra generated by $M_{\mu\nu}$ and
$\widetilde{M}_{\mu\nu}$ (Eqs. \eqref{37}, \eqref{38}, \eqref{551}
and \eqref{52}), but there are also other implementations of
Lorentz algebras compatible with the $\kappa$-Minkowski space-time
(Refs. \cite{klry08}, \cite{Bu-Kim}, \cite{Borowiec-3} and
\cite{kmps}). Further developments of this approach as well as its
applications to field theory, statistics, twist operators (see
Refs. \cite{Bu-Kim}, \cite{Govindarajan-1}, \cite{Govindarajan-2}
and \cite{Borowiec-3}) and dispersion relations \cite{bgmp10} will
be presented elsewhere.

\section*{Acknowledgements}
This work was supported by the Ministry of Science and Technology of the Republic of Croatia under contract No. 098-0000000-2865
and 177-0372794-2816.


\begin{thebibliography}{99}

\bibitem{Connes} A. Connes, \textit{Noncommutative Geometry}, Academic Press, 1994.

%%%%%%%%%%%%%%%%%%%%%%%%%%%%%%%%%%%%%%%%%%%%%%%%%%%%%%%%%%%%%%%%%%%%%%%%%%%%%%%%%%%%%%%%%%%%%%%%%%%%%%%%%%%%%%%%%%%%%%%%%%%%%%%%%%%%%%%%%%%%%

\bibitem{Majid-1} S. Majid, \textit{Foundations of Quantum Group Theory}, Cambridge University Press, 1995.

\bibitem{Majid-Ruegg} S. Majid and H. Ruegg, ``Bicrossproduct structure of $\kappa$-Poincar\'{e} group and noncommutative geometry'',
Phys. Lett. B {\textbf 334}, 348 (1994), hep-th/9404107.

\bibitem{Woronowicz} S. L. Woronowicz, ``Differential Calculus on Compact Matrix Pseudogroups (Quantum Groups)'',
Comm. Math. Phys. \textbf{122}, 125 (1989).

\bibitem{Aschieri} P. Aschieri, F. Lizzi and P. Vitale, ``Twisting all the way: from classical to quantum mechanics'',
Phys. Rev. D \textbf{77} 025037, (2008), arXiv:0708.3002v2.

\bibitem{Landi} G. Landi, \textit{An Introduction to Noncommutative Spaces and Their Geometries}, Lect. Notes Phys, \textbf{m 51} (1997),
hep-th/9701078v1.

\bibitem{Sitarz} A. Sitarz, ``Noncommutative differential calculus on the kappa-Minkowski space'',
Phys. Lett. B {\bf 349}, 42 (1995), hep-th/9409014.

\bibitem{Gonera} C. Gonera, P. Kosinski and P. Maslanka, ``Differential calculi on quantum Minkowski space'', J. Math. Phys. \textbf{37},
5820 (1996), arXiv:q-alg/9602007.

\bibitem{Dimitrijevic} M. Dimitrijevi{\'c}, L. M{\"o}ller and E. Tsouchnika, ``Derivatives, forms and vector fields on the
$\kappa$-deformed Euclidean space'', J. Phys. A: Math. Theor. {\textbf 37} (2004), hep-th/0404224.

\bibitem{Wess} J. Wess, ``Deformed coordinates spaces; Derivatives'', Lecture given at the Balkan workshop BW2003, August 2003,
Vrnja\v{c}ka Banja, Serbia, hep-th/0408080.

\bibitem{klry08} H. C. Kim, Y. Lee, C. Rim and J. H. Yee, ``Differential structure on the $\kappa$-Minkowski
spacetime from twist'', Phys. Lett. B {\textbf  671}, 398 (2009), arXiv:hep-th/0808.2866.

\bibitem{Bu-Kim} J.G. Bu, J.H. Yee and H.C. Kim, ``Differential structure on $\kappa$-Minkowski spacetime realized
as module of twisted Weyl algebra'', Phys. Lett. B \textbf{679}, 486 (2009), arXiv:0903.0040v2.

\bibitem{Meljanac-1} S. Meljanac and S. Kre\v{s}i\'{c}-Juri\'{c}, ``Noncommutative differential forms on the kappa-deformed space'',
J. Phys. A: Math. Theor. \textbf{42}, 365204 (2009), arXiv:0804.3072.

\bibitem{Meljanac-2} S. Meljanac and S. Kre\v{s}i\'{c}-Juri\'{c}, ``Differential  structure on $\kappa$-Minkowski space, and
$\kappa$-Poincar\'{e} algebra'', Int. J. Mod. Phys. A \textbf{26} (20), 3385 (2011), arXiv: 1004.4547.

\bibitem{Majid-2} S. Majid, ``Quantum Anomalies and Newtonian Gravity on Quantum Spacetime'', arXiv:1109.6190v1.

\bibitem{Lukierski-1} J. Lukierski, A. Nowicki, H. Ruegg and V. N. Tolstoy, ``Q-deformation of Poincar{\'e} algebra'', Phys. Lett. B
{\textbf 264}, 331 (1991).

\bibitem{Lukierski-2} J. Lukierski, A. Nowicki and  H. Ruegg, ``New  quantum Poincar{\'e} algebra,
and $\kappa$-deformed field theory'',  Phys. Lett. B {\textbf 293}, 344 (1992).

\bibitem{Lukierski-3} J. Lukierski and H. Ruegg, ``Quantum $\kappa$-Poincar{\'e} in any dimension'',
Phys. Lett. B {\textbf 329}, 189 (1994), hep-th/9310117.

\bibitem{Lukierski-4} J. Lukierski, M. Woronowicz, ``New Lie algebraic and quadratic
deformations of Minkowski space from twisted  Poincar{\'e} symmetries'', Phys. Lett. B
{\bf 633}, 116 (2006), hep-th/0508083.

\bibitem{Kosinksi} P. Kosi{\'n}ski and P. Ma{\'s}lanka, ``The duality between $\kappa$-Poincar{\'e}
algebra and $\kappa$-Poincar{\'e} group'', hep-th/9411033.

\bibitem{Zakrzewski} S. Zakrzewski, ``Quantum Poincar\'{e} group related to the kappa-Poincar\'{e} algebra'',
J. Phys. A \textbf{27}, 2075 (1994).

\bibitem{Amelino-Camelia} G. Amelino-Camelia  and M. Arzano, ``Coproduct and
star-product in field theories on Lie algebra noncomutative spacetime'', Phys. Rev.D  {\bf 65},  084044 (2002), hep-th/0105120.

\bibitem{Moller} L. M{\"o}ller, ``A symmetry invariant integral on $\kappa$-deformed spacetime'',
JHEP {\bf 0512}, 029 (2005), hep-th/0409128.

\bibitem{Agostini-2} A. Agostini,  G. Amelino-Camelia, M. Arzano, A. Marciano and R.A. Tacchi,
``Generalizing the Noether theorem for Hopf-algebra spacetime symmetries'', Mod. Phys. Lett. \textbf{A22} 1779-1786, (2007),
hep-th/0607221.


%%%%%%%%%%%%%%%%%%%%%%%%%%%%%%%%%%%%%%%%%%%%%%%%%%%%%%%%%%%%%%%%%%%%%%%%%%%%%%%%%%%%%%%%%%%%%%%%%%%%%%%%%%%%%%%%%%%%%%%%%%%%%%%%%%%%%%%%%%%%

\bibitem{Beggs} E. J. Beggs and S. Majid, ``Semiclassical differential structures'', Pac. J. Math. \textbf{224}, 1 (2006).


%%%%%%%%%%%%%%%%%%%%%%%%%%%%%%%%%%%%%%%%%%%%%%%%%%%%%%%%%%%%%%%%%%%%%%%%%%%%%%%%%%%%%%%%%%%%%%%%%%%%%%%%%%%%%%%%%%%%%%%%%%%%%%%%%%%%%%%%%%%

\bibitem{Amelino-Camelia-1} G. Amelino-Camelia, ``Testable scenario for relativity with minimum-length'', Phys. Lett. B \textbf{510},
255 (2001), hep-th/0012238.

\bibitem{Amelino-Camelia-2} G. Amelino-Camelia, ``Relativity in space-times with short-distance structure governed by an
observer-independent (Planckian) length scale'', Int. J. Mod. Phys. D {\textbf 11}, 35 (2002), gr-qc/0012051.

\bibitem{Amelino-Camelia-3} N. R. Bruno, G. Amelino-Camelia and J. Kowalski-Glikman, ``Deformed boost transformations
that saturate at the Planck scale'', Phys. Lett. B \textbf{522}, 133 (2001), hep-th/0107039.

\bibitem{Kowalski-Glikman-1} J. Kowalski-Glikman and S. Nowak, ``Double special relativity theories  as different
bases of  kappa-Poincar{\'e} algebra'', Phys. Lett. B \textbf{539}, 126 (2002), hep-th/0203040.

\bibitem{k-gn02} L. Freidel, J. Kowalski-Glikman and S. Nowak, ``Field theory on $\kappa$--Minkowski space revisited: Noether charges
 and breaking of Lorentz symmetry'', Int. J. Mod. Phys. A {\textbf 23}, 2687 (2008), arXiv:0706.3658.

\bibitem{Kowalski-Glikman-2} J. Kowalski-Glikman, ``Introduction to doubly special relativity'', Lect. Notes Phys.
\textbf{669}, 131 (2005), hep-th/0405273v1.

\bibitem{Kowalski-Glikman-3} J. Kowalski-Glikman and S. Nowak, ``Non-commutative space-time of doubly special relativity
theories'', Int. J. Mod. Phys. D \textbf{12}, 299 (2003).

\bibitem{Daskiewicz} M. Daskiewicz, K. Imilkowska, J. Kowalski-Glikman and S. Nowak, ``Scalar field theory on $\kappa$-Minkowski
space-time and doubly special relativity'', Int. J. Mod. Phys. A \textbf{20}, 4925 (2005).

\bibitem{dlw00} M. Daszkiewicz,  J. Lukierski and M. Woronowicz, ``Towards Quantum Noncommutative
$\kappa$-deformed Field Theory'', Phys. Rev. D \textbf{77}, 105007  (2008), arXiv:hep-th/0708.1561.

\bibitem{klm00} P. Kosinski, J. Lukierski and P. Maslanka, ``Local D=4 Field Theory on $\kappa$--Deformed Minkowski Space'', Phys. Rev. D \textbf{62}, 025004 (2000), arXiv:hep-th/9902037;

\bibitem{klry09} H. C. Kim, Y. Lee, C. Rim and J. H. Yee, ``Scalar field theory in $\kappa$-Minkowski spacetime from twist'', J. Math. Phys.
{\textbf 50}, 102304 (2009), arXiv:hep-th/0901.0049.

\bibitem{ms11} S. Meljanac and A. Samsarov, ``Scalar field theory on kappa-Minkowski spacetime and translation and Lorentz invariance'',
Int. J. Mod. Phys. A \textbf{26}, 1439 (2011), arXiv:1007.3943.

\bibitem{mstw11} S. Meljanac, A. Samsarov, J. Trampetic and M. Wohlgenannt, ``Scalar field propagation in the phi 4 kappa-Minkowski model'',
JHEP \textbf{1112}, 010 (2011), arXiv:1111.5553.

\bibitem{hjm11} E. Harikumar, T. Juric and S. Meljanac, ``Electrodynamics on $\kappa$-Minkowski space-time'',
Phys. Rev. D \textbf{84}, 085020 (2011), arXiv:1107.3936.

\bibitem{h00} E. Harikumar, ``Maxwell's equations on the $\kappa$-Minkowski spacetime and
Electric-Magnetic duality'', Europhys. Lett. \textbf{90}, 21001 (2010), arXiv:1002.3202v3.

\bibitem{dj11} M. Dimitrijevic and L. Jonke, ``A twisted look on kappa-Minkowski: U(1) gauge theory'', JHEP \textbf{1112},
080 (2011), arXiv:1107.3475.

\bibitem{Govindarajan-1} T. R. Govindarajan, K. S. Gupta, E. Harikumar, S. Meljanac and D. Meljanac, ``Twisted statistics in
$\kappa$-Minkowski spacetime'', Phys. Rev. D \textbf{77}, 105010 (2008), arXiv:0802.1576.

\bibitem{Govindarajan-2} T. R. Govindarajan, K. S. Gupta, E. Harikumar, S. Meljanac and D. Meljanac, ``Deformed osciallator
algebras and QFT in the $\kappa$-Minkowski spacetime'', Phys. Rev. D \textbf{80}, 025014 (2009), arXiv:0903.2355.

%%%%%%%%%%%%%%%%%%%%%%%%%%%%%%%%%%%%%%%%%%%%%%%%%%%%%%%%%%%%%%%%%%%%%%%%%%%%%%%%%%%%%%%%%%%%%%%%%%%%%%%%%%%%%%%%%%%%%%%%%%%%%%%%%%%%%%%


\bibitem{Meljanac-3} S. Meljanac, S. Kre\v{s}i\'{c}-Juri\'{c} and M. Stoji\'{c}, ``Covariant realizations of kappa-deformed space'',
Eur. Phys. J. C \textbf{51}, 229 (2007).

\bibitem{Meljanac-4} S. Meljanac and M. Stoji\'{c}, ``New realizations of Lie alegbra kappa-deformed Euclidean space'', Eur. Phys. J. C
\textbf{47}, 531 (2006).

\bibitem{ms06} S. Meljanac, A. Samsarov, M. Stoji\'{c} and K. S. Gupta, ``Kappa-Minkowski space-time and the star product
realizations'', Eur. Phys. J. C \textbf{53}, 295 (2008), arXiv:0705.2471.
% citirana svuda gdje i Meljanac-4

\bibitem{Borowiec-1} A. Borowiec and A. Pachol, ``$\kappa$-Minkowski Spacetimes and DSR Algebras: Fresh Look and Old Problems'',
SIGMA \textbf{6}, 086 (2010), arXiv:1005.4429.

\bibitem{Borowiec-2} A. Borowiec and A. Pachol, ``The classical basis for the $\kappa$-Poincar\'{e} Hopf algebra and doubly
special relativity theories'', J. Phys. A: Math. Theor. \textbf{43}, 045203 (2010), arXiv:0903.5251.

\bibitem{Borowiec-3} A. Boroweic and A. Pachol, ``$\kappa$-Minkowski spacetime as the result of Jordanian twist deformation'',
Phys. Rev. D \textbf{79}, 045012 (2009), arXiv:0812.0576.

\bibitem{kmps} D. Kova\v{c}evi\'{c}, S. Meljanac, A. Pacho{\l} and R. \v{S}trajn, ``Generalized $\kappa$-Poincar\'{e} algebras,
Hopf algebras and $\kappa$-Minkowski spacetime'', arXiv:1202.3305v1.

\bibitem{bgmp10} A. Borowiec, Kumar S. Gupta, S. Meljanac and A. Pacho{\l}, ``Constraints on the quantum gravity scale from
kappa--Minkowski spacetime'', Europhys. Lett. \textbf{92}, 20006 (2010), arXiv:0912.3299.





\end{thebibliography}
\end{document}